\documentclass[11pt]{article}
\pdfoutput=1
\usepackage[utf8]{inputenc}
\usepackage[T1]{fontenc}
\usepackage[normalem]{ulem}
\usepackage{verbatim}
\usepackage{amsmath,amssymb,amsthm,amsfonts,mathrsfs,dsfont,braket,array,float}
\usepackage{color,graphicx}
\usepackage{slashed}
\usepackage{enumitem}

\usepackage{epsfig}
\usepackage{mathabx}
\usepackage[mathscr]{eucal}
\usepackage{stmaryrd}
\usepackage{esint}
\usepackage{physics}
\usepackage{setspace}
\usepackage{braket}
\usepackage{float}
\usepackage{url}
\usepackage{mathtools}
\usepackage{bbold}
\usepackage{slashed}
\usepackage{tikz}
\usepackage{graphicx}
\usepackage{epstopdf}
\usepackage{subfigure}
\usepackage[labelfont=bf]{caption}
\usepackage{pgfplots}
\usepackage{fancybox}
\usepackage{eurosym}
\usepackage{tcolorbox}
\usepackage{tensor}
\allowdisplaybreaks

\usepackage[margin = 2.5cm]{geometry}
\usepackage[ragged]{footmisc}
\usepackage[bookmarks=true,bookmarksnumbered=false,
hyperindex=true,bookmarksopen=true,hyperfigures=true,
colorlinks=true,linkcolor=darkgray,citecolor=custom,urlcolor=custom,breaklinks]{hyperref}
\definecolor{webgreen}{rgb}{0, 0.5, 0}
\definecolor{webblue}{rgb}{0.06, 0.2, .65}
\definecolor{webblue2}{rgb}{0, 0.33, .71}
\definecolor{webred}{rgb}{0.9, 0.1, 0}
\definecolor{darkgreen}{rgb}{0,0.4,0.2}
\definecolor{custom}{rgb}{0.05,0.31,0.55}
\definecolor{darkgray}{rgb}{0.4,0.4,0.4}

\def\be{\begin{equation}}
	\def\ee{\end{equation}}

\numberwithin{equation}{section}

\def\mR{{\mathcal R}}

\usepackage{formatting}

\usepackage{shortcuts}

\pgfplotsset{compat=1.18} 

\begin{document} 
\begin{titlepage}
\begin{center}
\phantom{ }

{\bf \Large{
Tearing down spacetime with quantum disentanglement
}}
\vskip 0.4cm
Roberto Emparan${}^{1,2}$, Javier M. Mag\'an${}^{3}$
\vskip 0.2cm
\small{${}^{1}$ \textit{Departament de F\'isica Qu\`antica i Astrof\'isica, Institut de Ci\`encies del Cosmos}}
\vskip -.4cm
\small{\textit{  Universitat de Barcelona, Mart\'i i Franqu\`es 1, E-08028 Barcelona, Spain}}
\vskip -.25cm
\small{${}^{2}$  \textit{Institució Catalana de Recerca i Estudis Avançats (ICREA)}}
\vskip -.4cm
\small{\textit{Passeig Lluis Companys, 23, 08010 Barcelona, Spain}}
\vskip -.25cm
\small{  ${}^{3}$ \textit{Instituto Balseiro, Centro At\'omico Bariloche}}
\vskip -.4cm
\small{\textit{ 8400-S.C. de Bariloche, R\'io Negro, Argentina}}
\vskip 0.3cm
\href{mailto:emparan@ub.edu}{emparan@ub.edu}, 
\href{mailto:javier.magan@cab.cnea.gov.ar}{javier.magan@cab.cnea.gov.ar}

\begin{abstract}
A longstanding enigma within AdS/CFT concerns the entanglement entropy of holographic quantum fields in Rindler space. The vacuum of a quantum field in Minkowski spacetime can be viewed as an entangled thermofield double of two Rindler wedges at a temperature $T=1/2\pi$. We can gradually disentangle the state by lowering this temperature, and the entanglement entropy should vanish in the limit $T\to 0$ to the Boulware vacuum. However, holography yields a non-zero entanglement entropy at arbitrarily low $T$, since the bridge in the bulk between the two wedges retains a finite width. We show how this is resolved by bulk quantum effects of the same kind that affect the entropy of near-extremal black holes. Specifically, a Weyl transformation maps the holographic Boulware states to near-extremal hyperbolic black holes. A reduction to an effective two-dimensional theory captures the large quantum fluctuations in the geometry of the bridge, which bring down to zero the density of entangled states in the Boulware vacuum. Using another Weyl transformation, we construct unentangled Boulware states in de Sitter space.
\end{abstract}

\end{center}

\vfill

\end{titlepage}

\setcounter{tocdepth}{2}


\setlength{\parskip}{.2em}


\section{Introduction}\label{SecI}

If spacetime is holographically built up from the quantum entanglement of microscopic degrees of freedom \cite{Maldacena:2001kr,Ryu:2006bv,VanRaamsdonk:2010pw}, it should also be possible to split it apart by disentangling these same degrees of freedom. However, studying this phenomenon with holographic methods reveals a puzzle: the disentangled state appears to keep a large entanglement entropy \cite{Emparan:1999gf,Czech:2012be}. We review this problem below and then proceed to resolve it in the rest of the article. Interestingly, the solution involves bulk quantum effects of a kind brought to bear on another long-standing enigma in black hole thermodynamics, namely, the entropy of near-extremal Reissner-Nordstrom black holes \cite{Iliesiu:2020qvm}.

\paragraph{Entangled wedges.} A fundamental property of the Minkowski vacuum of a quantum field is that it can be described as an entangled state of the field theories on complementary Rindler wedges \cite{Unruh:1976db},
\begin{align}\label{TFD}
    |0\rangle_M=\frac1{\sqrt{Z}}\sum_i e^{-\pi E_i}|E_i\rangle_L |E_i\rangle_R\,.
\end{align}
Here $|E_i\rangle_{L,R}$ are eigenstates of the left and right Rindler (modular) Hamiltonians conjugate to the dimensionless time $t$ in the Rindler spacetime
\begin{align}\label{rindler}
    ds^2 = -\zeta^2 dt^2 +d\zeta^2 +d\mathbf{x}^2\,.
\end{align}
The state \eqref{TFD} can be viewed as a thermofield-double state built out of excitations at temperature $T=1/2\pi$ above the Boulware vacuum $|0\rangle_L |0\rangle_R$.\footnote{The most appropriate language here is that of algebraic quantum field theory, since the Hilbert space does not factorize into left and right in the continuum limit. Factorized ``split states'' exist (from the split property  \cite{Doplicher:1984zz,haag2012local}) when we allow a regularizing space between the left and right sides. But we may just imagine discretizing the geometry, and the puzzles that we describe below will still appear. } The entanglement entropy of this state is infinite for two reasons. The first is simply the infinite extent of the plane $\{\mathbf{x}\}$, but we can compactify it to finite size and factor out its area $A_{n-2}$. The resulting entropy density is still infinite for a more important reason: modes of arbitrarily short wavelength are entangled across the divide $\zeta=0$. If we introduce a small length cutoff $\varepsilon$, then the leading divergent term takes the form \cite{Bombelli:1986rw,Srednicki:1993im}
\begin{align}\label{Sdiv}
    S=\frac{A_{n-2}}{\varepsilon^{n-2}}\, s\,.
\end{align}
The dimensionless finite quantity $s$ can be regarded as a local measure of entanglement. Since its precise value varies with the choice of cutoff, it is often disregarded as lacking universality. However, if we introduce a physical cutoff, the mere fact that $s$ is not zero is physically significant: there is a large entanglement between $L$ and $R$. The problem we will describe is, in this sense, universal. It is also present, although less dramatically, for the smaller contributions to the entanglement entropy that diverge logarithmically with a coefficient that is cutoff-independent. We briefly discuss these different quantities for free theories in appendix~\ref{app:free}.

One can also construct entangled states at different temperatures $T=1/\beta$, 
\begin{align}\label{psibeta}
    |\Psi_\beta\rangle=\frac1{\sqrt{Z}}\sum_i e^{-\beta E_i/2}|E_i\rangle_L |E_i\rangle_R\,.
\end{align}
We are interested in lowering the entanglement temperature, taking $\beta> 2\pi$. Unlike the inertial vacuum \eqref{TFD}, these states are singular on the Rindler horizon $\zeta=0$: the stress tensor diverges due to the infinite forces required to prevent the field from freely falling across the horizon. The entanglement entropy still diverges like \eqref{Sdiv},
\begin{align}\label{Sdivbeta}
    S(\beta)=\frac{A_{n-2}}{\varepsilon^{n-2}} \, s(\beta)\,,
\end{align}
but we expect that $s(\beta)$ will decrease as we lower $T$ (keeping the cutoff fixed in the manner explained below), and approach zero as $T\to 0$. In this limit, the states approach the Boulware vacuum, $|\Psi_{\beta\to\infty}\rangle\to |0\rangle_L |0\rangle_R$, which is an unentangled product state. That is, as long as the vacuum of the modular Hamiltonian is non-degenerate (and without (super)symmetry protection, large degeneracies are not expected), the limit $T\rightarrow 0$ will erase the entanglement between the two sides. 

\paragraph{Entangling spacetime.} The previous considerations apply to any local quantum field theory, even interacting ones. Strongly coupled holographic CFTs in Rindler space are interesting since one can argue that large regions of the dual AdS geometry emerge as a consequence of the entanglement between the CFT$_L$ and CFT$_R$ \cite{VanRaamsdonk:2010pw,Czech:2012be}. To see how this happens, we need the bulk dual of the CFT state \eqref{TFD} with the boundary geometry \eqref{rindler}. It is easy to see that this is simply the empty AdS$_{n+1}$ spacetime written in accelerated coordinates (see appendix~\ref{app:coords}),
\begin{align}
    ds^2&= -\left( r^2-\ell^2\right)  dt^2+\frac{\ell^2 dr^2}{ r^2-\ell^2} + r^2\, \frac{d\zeta^2 +d\mathbf{x}^2}{\zeta^2}\nonumber\\
    &=\frac{r^2}{\zeta^2}\left[ -\left(1-\frac{\ell^2}{r^2}\right) \zeta^2 dt^2 +d\zeta^2 +d\mathbf{x}^2 \right]+\frac{\ell^2}{r^2} \frac{dr^2}{1-\frac{\ell^2}{r^2}}\,.\label{rindlerads}
\end{align}
In the first line we present it in a form resembling a black hole with a horizon at $r=\ell$ that spans the hyperboloid $H_{n-1}$ with coordinates $(\zeta,\mathbf{x})$. In the second line we have pulled out a factor $r^2/\zeta^2$ multiplying the square brackets, so that the appearance of the Rindler spacetime \eqref{rindler} at the conformal asymptotic boundary $r\to\infty$ becomes manifest. With the above normalization for the timelike Killing vector $\partial_t$, the temperature of the horizon is $1/2\pi$. Then the spacetime \eqref{rindlerads} describes the CFT state \eqref{TFD}, which is globally regular \cite{Emparan:1999gf,Casini:2011kv}. Ref.~\cite{Czech:2012be} argued that the regions of the bulk AdS spacetime beyond the acceleration horizons, which cannot be reconstructed from the states in CFT$_L$ and CFT$_R$ separately, must be thought of as emerging from the entanglement between them. 

The entanglement entropy density $s$ of this state can be obtained using the holographic RT prescription \cite{Ryu:2006bv}: on a constant $t$ section in the bulk, we seek a minimal surface that is anchored on the plane $\zeta=0$ at the boundary. This surface is the horizon at $r=\ell$. The ultraviolet divergence of the entropy near the Rindler horizon at $\zeta=0$ corresponds to the infinite volume of the hyperboloid $H_{n-1}$. We regularize this volume by compactifying $\{\mathbf{x}\}$ and restricting to $\zeta\geq \varepsilon$. We find an entropy of the form of \eqref{Sdiv}, with 
\begin{align}\label{sads}
    s=\frac{\ell^{n-1}}{(n-2)4G}\,.
\end{align}
The interpretation is that this entropy, which measures the left-right entanglement in the state \eqref{TFD}, is manifested geometrically in the `emergent' bridge in the bulk between the two sides. 

This calculation is the holographic realization of the fact that the entanglement entropy of a CFT in Rindler space is the same as its thermal entropy in the hyperboloid. Two comments are in order: (i) The regularization of Rindler divergences through a map to a hyperbolic volume works in exactly the same way for any conformal field theory, holographic or not. (ii) The holographic entropy in $n=4$ exactly reproduces the weak coupling result for the large $N$ limit of $\mathcal{N}=4$ SYM theory, with proper care for spin-one fields \cite{Eling:2013aqa}.

\paragraph{Disentangling spacetime.} 
We now want to see how the geometric connection between the left and right sides diminishes as we disentangle the quantum state. For this purpose, we examine the bulk duals of the states \eqref{psibeta} as we lower the temperature. These duals are not obtained by choosing an imaginary time period $\beta> 2\pi$  in \eqref{rindlerads}, since this would create a Euclidean conical singularity in the bulk. Instead, the required solutions are the hyperbolic (a.k.a.~`topological') AdS black holes that were given an AdS/CFT interpretation in \cite{Emparan:1999gf,Czech:2012be}. In $n+1$ dimensions their geometry can be written as
\begin{align}\label{hyperbh}
    ds^2=\frac{r^2}{\zeta^2}\left( -f(r) \zeta^2 dt^2 +d\zeta^2 +d\mathbf{x}^2 \right)+\frac{\ell^2}{r^2} \frac{dr^2}{f(r)}\,,
\end{align}
with
\begin{align}\label{fofr}
    f(r)=1-\frac{\mu}{r^n}-\frac{\ell^2}{r^2}\,.
\end{align}
When $\mu=0$ this is the same metric as \eqref{rindlerads}, i.e., locally the same as AdS spacetime. When $\mu\neq 0$ the curvature is not constant and we have actual black holes with singularities in their interiors. We see again the Rindler geometry \eqref{rindler} at the conformal boundary, but now the temperature of the state, i.e., the temperature of the horizon, is not $T=1/2\pi$ (the details will be given later). These are the bulk duals of the CFT states $|\Psi_\beta\rangle$. We are interested in those with $\mu<0$, whose horizons have $T<1/2\pi$ and thus are less entangled than \eqref{TFD}.\footnote{A different class of `partially entangled thermal states' of the kind constructed in \cite{Goel:2018ubv,Balasubramanian:2022gmo,Balasubramanian:2022lnw,Antonini:2023hdh} can also be studied in this setting.} An interesting feature is that, even though these CFT states are singular, and indeed their stress tensor diverges at $\zeta=0$ \cite{Emparan:1999gf,Czech:2012be}, the bulk geometries are regular everywhere on and outside the black hole horizon. 

Despite being neutral, the hyperbolic black holes with $\mu<0$ resemble in many respects Reissner-Nordstrom black holes. They exist with regular horizons (inner and outer) down to a minimum $\mu=\mu_0<0$, where the black hole becomes extremal with $T=0$. Ref.~\cite{Czech:2012be} found that they reproduce several of the expected properties of the states $|\Psi_\beta\rangle$. For instance, the geodesic distance between the two asymptotic regions increases as $\mu$ is lowered. This is a signal that the correlations between the left and right CFTs become weaker as $T$ decreases. In the bulk, the Einstein-Rosen bridge between the two sides becomes a very long throat at small $T$. Furthermore, the entanglement entropy density $s$, measured from the area of the horizon, also becomes smaller as we lower $T$: the throat narrows down. 

Although these features work as expected, on closer examination a sharp puzzle appears: when $T\to 0$ the entanglement entropy does not approach zero. Instead one finds
\begin{align}\label{sext}
    s(\beta\to\infty)\to s_0=\left(\frac{n-2}{n}\right)^{\frac{n-1}{2}}\frac{\ell^{n-1}}{(n-2)4G}\neq 0\,.
\end{align}
This is indeed a smaller entropy than \eqref{sads}, but the fact that $s$ has a lower non-zero bound is wholly unexpected. It says that as we approach the state dual to this black hole, namely the Boulware vacuum $|0\rangle_L |0\rangle_R$, we retain a large  $O(1/G)$ amount of entanglement---when it should shrink towards zero. 

\paragraph{Quantum disassembling.} The way out of this contradiction becomes clear when we relate it to a recently solved problem: the $O(1/G)$ entropy of near-extremal non-supersymmetric black holes \cite{Iliesiu:2020qvm}. The resolution of the latter lies in the existence of large quantum fluctuations near the horizon at low temperatures, which are described by an effective one-dimensional Schwarzian theory \cite{Maldacena:2016upp,Stanford:2017thb,Bagrets:2016cdf,Mertens:2017mtv,Kitaev:2018wpr,Yang:2018gdb}. The quantum one-loop effects bring down the low-temperature entropy, with the effect that the density of states vanishes as the energy above extremality approaches zero. These quantum effects near extremality have been studied recently in several scenarios \cite{Ghosh:2019rcj,Heydeman:2020hhw,Boruch:2022tno,Iliesiu:2022kny,Iliesiu:2022onk,Banerjee:2023quv,Kapec:2023ruw,Rakic:2023vhv,Banerjee:2023gll}.

In the remainder of this article we will explain how the hyperbolic black holes develop an AdS$_2$ throat close to extremality, with dynamics captured by an effective two-dimensional JT theory. Then, properly accounting for quantum effects will reduce the entanglement entropy of the low-temperature thermofield Rindler states below the naive result \eqref{sext}. In particular, we will show that the modular density of entangled states vanishes when the Rindler energy goes to zero. A simple modification of the construction allows us to disentangle the state of a CFT in de Sitter space and obtain the Boulware-de Sitter vacuum.

The entanglement of a two-dimensional holographic CFT, corresponding to $n=2$, shows qualitative differences compared to $n>2$ and is discussed in appendix~\ref{app:2d}.

\bigskip 

\noindent\textit{Note on modular units.} The interpretation in Rindler space makes it natural to define $t$ , $E$ and $T$ as the dimensionless modular time, energy and temperature. We will also keep them dimensionless when we discuss black holes in AdS as in \eqref{hyperbh}. Conventional units are restored by substituting $t\to t/\ell$, $E\to \ell E$, $T\to \ell T$, where $\ell$ is the AdS radius.

\section{Hyperbolic black holes near extremality}\label{sec:hyperbh}

The hyperbolic black holes \eqref{hyperbh} have an event horizon at the largest real root $r_+$ of $f(r)$, which solves the equation
\begin{align}
    \mu=r_+^n\left( 1-\frac{\ell^2}{r_+^2}\right)\,.
\end{align}
It is often convenient to regard this expression as giving $\mu$ in terms of the parameter $r_+$, rather than the other way around. We see that $\mu$ is negative when $r_+<\ell$. There exists a range of parameters where such solutions are black holes with regular horizons. To see this, observe that the temperature
\begin{align}
    T=\frac{n r_+^2-(n-2)\ell^2}{4\pi \ell r_+}\;,
\end{align}
is non-negative as long as 
\begin{align}\label{r0mu0}
    r_+\geq r_0\equiv\sqrt{\frac{n-2}{n}}\,\ell\,,\quad \textrm{i.e.,}\quad \mu\geq \mu_0\equiv-\frac{2}{n}\left(\frac{n-2}{n}\right)^{(n-2)/2}\ell^n\,.
\end{align}
For the metrics \eqref{hyperbh}, this guarantees that the horizon at $r=r_+$ is smooth. The temperature decreases monotonically from $T=1/2\pi$ for $\mu=0$, down to the extremal limit $T=0$ for $\mu= \mu_0$. Hyperbolic black holes in this range have the same causal structure as Reissner-Nordstrom-AdS black holes \cite{Emparan:1999gf}. The required repulsion is not provided by an electric field, but rather by the hyperbolic negative curvature. This will become apparent in the next section.

Hyperbolic black holes are sometimes called `topological black holes' because, by taking discrete quotients of $H_{n-1}$, their horizon can be made into a compact space of non-trivial topology. For instance, in $n=3$ one can obtain surfaces of arbitrary genus $g>1$. This compactification renders finite the volume $V_H$ of $H_{n-1}$ and hence regularizes the entropy of the black hole. However, in this article we will not do this since it is not well motivated by the Rindler space interpretation. Instead, we take
\begin{align}\label{VHreg}
    V_H&=\int d^{n-2}x\,\, d\zeta\, \zeta^{1-n}=A_{n-2}\int_{\varepsilon}^\infty d\zeta\,\zeta^{1-n}\nn\\
    &=\frac1{n-2}\frac{A_{n-2}}{\varepsilon^{n-2}}\,.
\end{align}
The energy of these black holes is 
\begin{align}\label{Ebh}
    E=(n-1)\frac{V_H}{16\pi G}\frac{\mu}{\ell}\,.
\end{align}
We have ignored the Casimir energy of the CFT vacuum in $H_{n-1}$ when $n$ is even  \cite{Emparan:1999gf}, since it has no relevance for us. More important is the extremal black hole energy for $\mu=\mu_0$,
\begin{align}
    E_0=&-\frac{n-1}{n}\left(\frac{n-2}{n}\right)^{\frac{n-2}{2}}\frac{\ell^{n-1}V_H }{8\pi G}\,.
\end{align}
When interpreting the system as the dual of the CFT in Rindler space, we would normally subtract this as a ground state energy, and thus set to zero the Boulware vacuum energy. This could be done with an appropriate counterterm. 

The energy and entropy diverge due to the factor $V_H$, which we always consider regularized as in \eqref{VHreg}. After factoring it out, the Bekenstein-Hawking entropy of the horizon is of the form \eqref{Sdivbeta} with
\begin{align}\label{sbh}
    s=\frac{r_+^{n-1}}{(n-2)4G}\,.
\end{align}
This is interpreted as the entanglement entropy of the states $|\Psi_\beta\rangle$ in the boundary theory. For the non-singular entangled state $|\Psi_{2\pi}\rangle$, eq.~\eqref{TFD}, with $\mu=0$, we recover the entanglement entropy \eqref{sads}. For the unentangled Boulware vacuum $|\Psi_{\infty}\rangle$ with $\mu=\mu_0$, it implies the troublesome non-zero result \eqref{sext}, that is,
\begin{align}\label{S0}
    S_0=\left(\frac{n-2}{n}\right)^{\frac{n-1}{2}}\frac{\ell^{n-1}V_H}{4G}\,.
\end{align} 

To go near extremality and near the horizon, we take
\begin{align}
    r_+=r_0+\rho_+\,,\qquad r=r_0+\rho\,,
\end{align}
and expand in small $\rho_+,\rho\ll r_0$, both of the same order. The metric \eqref{hyperbh} becomes
\begin{align}\label{ads2H}
    ds^2=-n(\rho^2-\rho_+^2)dt^2+\frac{\ell^2}{n}\frac{d\rho^2}{\rho^2-\rho_+^2}+(r_0+\rho)^2\,\frac{d\zeta^2+d\mathbf{x}^2}{\zeta^2}\,.
\end{align}
This is the product of thermal AdS$_2$ with radius 
\begin{align}
    L_2=\frac{\ell}{\sqrt{n}}=\frac{r_0}{\sqrt{n-2}}\,,
\end{align}
times a hyperboloid $H_{n-1}$ of almost constant radius $r_0$. The fluctuations in the size of $H_{n-1}$ have been retained because they will dominate the dynamics at low temperatures. We can eliminate $\rho_+$ in favor of $T$ using that in this limit
\begin{align}
    T=\frac{n}{2\pi \ell}\,\rho_+\,.
\end{align}

Observe that these are all large AdS black holes with $r_+\lesssim \ell$, and the radii of the AdS$_2$ and $H_{n-1}$ factors are always of the same parametric order $\sim \ell$. In contrast, the RN-AdS solutions have one more parameter, and this allows to separate the sizes of the two factors in the geometry\footnote{The scale separation $R_{\textrm{AdS}_2}\ll R_{H_{n-1}}$ can be obtained if $n$ is regarded as a large parameter.}.  Nevertheless, to smoothly connect the throat near the horizon \eqref{ads2H} to the outer zone it is enough to be in a low-temperature regime where $T\ll 1$, i.e., $\rho_+\ll \ell,r_0$. If we define the outer zone as the region $r-r_0\gg r_+-r_0=\rho_+$, where the metric is well approximated by the extremal solution, then it overlaps with the throat at radii $r$ such that $\rho_+\ll r-r_0\ll r_0$.

Close to extremality, the energy \eqref{Ebh} and entropy \eqref{sbh} behave as
\begin{align}
    E(T)=&E_0+\frac{2\pi^2}{M_b} T^2 +O\left(T^3\right)\,,\label{EofT}\\
    S(T)=&S_0+\frac{4\pi^2}{M_b} T+O\left(T^2\right)\,.\label{SofT}
\end{align}
Here 
\begin{align}\label{massgap}
    M_b^{-1}=|E_0|=\frac{n-1}{n}\left(\frac{n-2}{n}\right)^{\frac{n-2}{2}}\frac{\ell^{n-1}V_H}{8\pi G}\;,
\end{align}
is the characteristic mass scale for excitations above extremality, which appears due to the quadratic dependence on $T$ \cite{Preskill:1991tb}. It would vanish if we let $V_H\to\infty$, so we will keep the regulator $\varepsilon$ small but non-zero.

Conformal field theories in the hyperboloid have a well-known instability due to the coupling of conformal scalars to the negative curvature, which gives them a tachyonic potential unbounded below. In the holographic dual, the hyperbolic black hole is unstable to the spontaneous nucleation of branes, at least if there exist BPS branes (see \cite{Barbon:2010us,Buchel:2004rr}). It may then happen that the dominant configuration has many branes and is a smaller, very non-classical spacetime, and our analysis indicates that the latter is the case. The instability is suppressed when $T\geq 1/2\pi$, since the energy added to the black hole attracts the branes and opposes the effective repulsion from the curvature. In dual terms, the thermal energy lifts the tachyon (adding suitable mass terms can also stabilize the theory \cite{Buchel:2004rr}). However, we are interested in very low temperatures. The nucleation rate of the branes is exponentially suppressed when $V_H$ is very large, but, as we have seen, in our analysis we keep the regulator small but non-zero, so issues may remain. We will not dwell on this question any longer, since the large quantum effects that we will find seem to require the revision of this instability. Other consequences of the divergent volume of $H_{n-1}$ related to the points above will be discussed later.

\section{Quantum throat dynamics}\label{sec:quthroat}

The value \eqref{SofT} accounts for the leading contribution to the entropy from the semiclassical saddle points \eqref{hyperbh} of the gravitational path integral. It neglects the possibility of large quantum effects at low $T\ll 1$. We now show that these effects drastically modify the result.

Following \cite{Nayak:2018qej,Iliesiu:2020qvm}, we study the corrections in the throat \eqref{ads2H} by dimensionally reducing on the hyperboloid to obtain an effective two-dimensional theory. This is sensible, because even though the hyperboloid volume is infinite (or arbitrarily large, when $\zeta$ is cut off), the finite curvature radius introduces an $O(1)$ gap in the spectrum of the Laplacian and therefore at temperature $T\ll 1$ the lowest modes dominate the dynamics. 

To drive our main point home more clearly, we will begin by focusing on fluctuations that are homogeneous in the hyperbolic space. Later we will include the zero-mode fluctuations that break homogeneity, and see that they do not alter our conclusions. 

\subsection{Homogeneous zero modes}

Starting from the Einstein-Anti de Sitter Euclidean action, 
\be 
I=-\frac{1}{16\pi G}\int_{\mathcal{M}}d^{n+1}x \sqrt{g_{n+1}}\left( R_{n+1}+\frac{n(n-1)}{\ell^2}\right )-\frac{1}{8\pi G}\int_{\partial\mathcal{M}}d^{n}x \sqrt{h_n}\,K_n\,,
\ee
we consider geometries of the form
\begin{align}\label{homog}
    ds^2=\Phi^{-\frac{n-2}{n-1}}g_{\mu\nu}dx^\mu dx^\nu +r_0^2\,\Phi^{\frac2{n-1}}\,dH_{n-1}\,.
\end{align}
The two-dimensional metric $g_{\mu\nu}$ and the dilaton $\Phi$ depend only on the coordinates $x^\mu=(\tau,r)$. The $H_{n-1}$ factor is the unit-radius hyperboloid, and $r_0$ is for now a fiducial length scale. It will later correspond to the extremal horizon radius. With this ansatz, the action reduces to a two-dimensional dilaton gravity theory
\begin{align}\label{dilgra}
    I=-\frac{V_H r_0^{n-1}}{16\pi G}\int d^2x \sqrt{g}\left[ \Phi R - 2U(\Phi)\right] -\frac{V_H r_0^{n-1}}{8\pi G}\int dx \sqrt{h}\,\Phi K\;,
\end{align}
with potential
\begin{align}
    U(\Phi)=\frac{(n-1)(n-2)}{2r_0^2}\Phi^{-\frac1{n-1}}-\frac{n(n-1)}{2\ell^2}\Phi^{\frac1{n-1}}\,.
\end{align}
The two terms in $U(\Phi)$---the first from the reduction on the hyperboloid, the second from the higher-dimensional cosmological constant---have opposite signs and thus can balance each other. This balance permits the existence of a near-extremal regime without any charge. Before examining this limit, let us note that Birkhoff's theorem allows us to obtain the complete solution to the classical theory, which is
\begin{align}\label{sol1}
    ds^2_2=\Phi^\frac{n-2}{n-1}\left( \ell^2 F(r) d\tau^2+\frac{dr^2}{F(r)}\right)\,,\qquad \Phi(r)=\left(\frac{r}{r_0}\right)^{n-1}\,,
\end{align}
where (see \cite{Grumiller:2002nm}) 
\begin{align}\label{sol2}
    F(\Phi)&=-\Phi^{-\frac{n-2}{n-1}}\left( k+\frac{2r_0^2}{(n-1)^2}\int^\Phi d\tilde\Phi\, U(\tilde\Phi)\right)\nn\\
    &=-1-k\,\Phi^{-\frac{n-2}{n-1}}+\frac{r_0^2}{\ell^2}\,\Phi^\frac2{n-1}\;,
\end{align}
with integration constant $k$. Setting $k=\mu/(r_0^{n-2}\ell^2)$ reproduces the hyperbolic black hole \eqref{hyperbh}.

Extremal solutions appear when $F(\Phi)$ and $F'(\Phi)$ have simultaneous zeros. Therefore,
to go near extremality and close to the horizon, we zoom in near the zeroes of $U(\Phi)$. We can normalize $\Phi$ so that $U(\Phi=1)=0$ and $r_0$ is the extremal horizon radius, which is then fixed to
\begin{align}
    r_0=\sqrt\frac{n-2}{n} \,\ell\,,
\end{align}
i.e., we recover \eqref{r0mu0}.

To separate the dynamics of the throat from the region outside it, we introduce a curve at a fixed value of $\Phi$, with fixed intrinsic metric and fixed proper length $l_c$. We set it at a radius $r=r_b=r_0+\delta r_b$,
so the proper length of the boundary curve in the extremal black hole geometry is
\begin{align}
    l_c=\beta\ell\sqrt{F(r_b)}=\beta \ell\,\frac{\delta r_b}{L_2} = \beta \sqrt{n}\,\delta r_b\,,
\end{align}
where $L_2$ is the AdS$_2$ radius that we found in \eqref{ads2H}.
Near the horizon we set
\begin{align}\label{Phiphi}
    \Phi=\frac{8\pi G}{V_H r_0^{n-1}}\left( \phi_0 +\phi(r)\right)\,,\qquad \phi(r)\ll\phi_0\equiv \frac{V_H r_0^{n-1}}{8\pi G}\,.
\end{align}
It is customary to use, instead of $\delta r_b$, a cutoff $\epsilon$ near the mouth of the throat, such that $\epsilon=L_2^2/(\ell \delta r_b)$,
and then the dilaton at the boundary curve is 
\begin{align}
    \phi(r_b)=\frac{\phi_b}{\epsilon}\,,\qquad 
    \phi_b 
    = M_{b}^{-1}\;.
\end{align}
Here $M_{b}$ is the mass scale that we identified from the thermodynamics in \eqref{massgap}. It was formerly referred to as the mass gap, but is more properly viewed as the scale of $SL(2,\mathbb{R})$ symmetry breaking in the AdS$_2$ throat \cite{Iliesiu:2020qvm}. Observe that even if $\delta r_b/\ell$ and $\epsilon$ are parametrically $O(1)$, the curve length $l_c/\ell$ and $\phi_b$ are both very large for large black holes close to extremality.

Plugging \eqref{Phiphi} in \eqref{dilgra}, the action for the region near the horizon takes the JT form
\be\label{Inear}
I_\textrm{near}=-\frac12\int_{M_\textrm{near}} d^{2}x \sqrt{g}\left[\phi_0 R+\phi \left( R +\frac{2}{L_2^2}\right)+\mathcal{O}\left(\frac{\phi}{\phi_0}\right)^2 \right] \,.
\ee
To this, we must add the action of the outer region. The geometry there is very approximately the extremal black hole metric, and away from the throat the dilaton is large and quantum fluctuations are comparatively small. This means that the contribution to the action from the bulk of the outer region can be computed on-shell in the extremal geometry. Using counterterm subtraction (and neglecting the Casimir energy, if present) this bulk action gives $\beta E_0$ from the mass of the extremal black hole. There only remains to include a boundary term for the fluctuations of the surface $\partial M_\textrm{near}$ that separates this region from the throat. A straightforward calculation then results in
\begin{align}
\label{j0}
I =&\;\beta E_0-\frac12\int_{M_\textrm{near}} d^{2}x \sqrt{g}\left[\phi_0 R+\phi \left( R +\frac{2}{L_2^2}\right)+\mathcal{O}\left(\frac{\phi}{\phi_0}\right)^2 \right]\nonumber\\ &-\int_{\partial M_\textrm{near}} du\, \sqrt{h}\left[\phi_0 K+\frac{\phi_b}{\epsilon} \left( K -\frac{1}{L_2}\right)\right]\,.
\end{align}
The action of JT gravity is now supplemented with the correct boundary terms, and from this point on the procedure is well known  \cite{Maldacena:2016upp,Mertens:2022irh}. The dilaton can be exactly integrated out enforcing $R=-2/L_2^2$, the topological terms in \eqref{j0} give the classical extremal entropy $S_0$ \eqref{S0}, and the only dynamics comes from the extrinsic curvature $K$, which yields the Schwarzian theory of boundary reparametrizations $\tau(u)$ from the broken $SL(2,\mathbb{R})$ symmetry, namely,
\begin{align}
\label{j1}
I 
&=\beta E_0-S_0-\phi_b\int_0^\beta du\, \textrm{Sch}(\tau,u)\,.
\end{align}
If we evaluate it on the classical solution at temperature $\beta^{-1}$ we obtain
\begin{align}\label{schwsol}
    \phi_b\int_0^\beta du\, \textrm{Sch}\left(\tan \frac{\pi u}{\beta},u\right)=\frac{2\pi^2 \phi_b}{\beta}=\frac{2\pi^2 T}{M_b}\,,
\end{align}
which reproduces the leading term in the free energy above extremality \eqref{EofT}, \eqref{SofT}. However, when $\beta\phi_b> 1$ the theory is strongly coupled and quantum effects become important. Fortunately, this theory can be quantized in several ways \cite{Stanford:2017thb,Bagrets:2016cdf,Mertens:2017mtv,Kitaev:2018wpr,Yang:2018gdb} to give a one-loop-exact partition function 
\be \label{schwZ}
Z=
e^{S_0-\beta E_0+\frac{2\pi^2 \phi_b}{\beta}}\,\left(\frac{\phi_b}{\beta}\right)^{3/2}\;.
\ee
The negative energy $E_0<0$ might look problematic, but for the purposes of Rindler interpretation we can subtract it as the ground state energy. The prefactor $\propto \;\beta^{-3/2}$ accounts for the quantum fluctuations, which strongly suppress the partition function at low temperatures $\beta\gg 1$. They modify the energy and entropy \eqref{EofT}, \eqref{SofT} as
\bea \label{qSE}
E(T)&=&E_0+\frac{2\pi^2}{M_b} T^2+\frac{3}{2}T+O(T^3)\;,\nonumber\\
S(T)&=&S_0+ \frac{4\pi^2}{M_b} T+\frac32\left(1+\log \frac{T}{M_b}\right)+O(T^2)\;.
\eea
We see that as $T\rightarrow 0$ the quantum log term drastically reduces the entropy from its semiclassical value $S_0$. We have been motivated to interpret this entropy as a measure of the entanglement between two Rindler wedges. We then obtain the result we sought: at entangling temperatures $T\ll 1$, the entanglement entropy between the left and right CFTs decreases to values much smaller than the $O(1/G)$ leading semiclassical result.

Crucially, this entanglement entropy does not arise from a semiclassical saddle point with small quantum fluctuations, as would be captured by a quantum-corrected RT formula still reliant on a classical notion of geometry \cite{Faulkner:2013ana,Engelhardt:2014gca}. In the path integral approach to the entanglement entropy \cite{Lewkowycz:2013nqa}, there are large quantum fluctuations around the saddle point \eqref{hyperbh} at low $T$. 

The result for the entropy is somewhat muddled by the logarithmic divergence as $T\to 0$. A better-behaved quantity is the modular density of entangled states with energy $E$. It can be extracted from \eqref{schwZ} to give
\be 
\rho(E)=e^{S_0}\sinh \left(2\pi \sqrt{2\phi_b(E-E_0)}\right)\,\Theta(E-E_0)\;,
\ee
and we see that it vanishes as the Boulware vacuum at $E\rightarrow E_0$ is approached. This is our main conclusion. It will remain qualitatively valid after we complete the analysis in the next subsection.

\subsection{Inhomogeneous zero modes}
\label{appzeromodes}

The result \eqref{schwZ} includes the quantum fluctuations of the throat that preserve the homogeneity of the hyperboloid. However, we must also account for zero modes of inhomogeneous fluctuations. To this end, we generalize the ansatz \eqref{homog} to 
\be
   ds^2=\Phi^{-\frac{n-2}{n-1}}g_{\mu\nu}dx^\mu dx^\nu +r_0^2\,\Phi^{\frac2{n-1}}\,h_{mn} (dy^m +\mathbf{A}^a\xi_a^m)(dy^n +\mathbf{A}^b\xi_b^n)\;,
\ee
where $\mathbf{A}^a=A_\mu^a(x)dx^\mu$ is a non-abelian gauge field in the adjoint representation of the group $SO(1,n-1)$ of isometries of the hyperbolic space $H_{n-1}$, which are generated by the vector fields $\xi_a$. In the dimensional reduction of the Einstein-AdS$_{n+1}$ action, these fields add to the dilaton gravity theory \eqref{dilgra} a Yang-Mills action
\be 
I_{\mathbf{A}}=-\frac{1}{2g_{YM}^2}\int d^2x \sqrt{g}\, \Phi^{\frac{n+2}{2}} \,\textrm{Tr} \left(F_{\mu\nu}F^{\mu\nu}\right)\;,
\ee
where $F=dA-A\wedge A$ and we have defined the gauge coupling to be
\be 
g_{YM}^2\equiv\frac{n(n-1)4\pi G}{V_H r_0^{n+1}}=\frac{n(n-1)}{2r_0^2\phi_0}\;.
\ee
This coupling would vanish in the non-compact limit $V_H\to \infty$, which is, again, a reason why we work with a small but finite cutoff $\varepsilon$.

Two-dimensional gauge theories of this type have been extensively studied in the literature \cite{Migdal:1975zf,Migdal:1975zg,Migdal:1983qrz,Rusakov:1990rs,Fine:1990zz,Fine:1991ux,Blau:1991mp,Witten:1991we,Witten:1992xu,Cordes:1994fc,Ganor:1994bq,Tseytfin_1995,Constantinidis:2008ty,Iliesiu:2019xuh,Iliesiu:2019lfc,Kapec:2019ecr}, most often for compact gauge groups. The particularities associated with compact groups include that the spectrum of irreducible representations (which becomes the spectrum of the Hamiltonian) is discrete, the irreps have finite dimensions, and the Casimirs are positive. Considerations of non-compact gauge groups, as we find here, have also appeared \cite{Tseytfin_1995,Constantinidis:2008ty,Iliesiu:2019xuh,Iliesiu:2019lfc}. 

The quantization of these theories has been solved in \cite{Iliesiu:2019xuh,Iliesiu:2019lfc} and we can directly borrow from them to go to the main results. Fixing the holonomy $c$ of the gauge field at the boundary, the partition function of the theory in the disk can be expressed in terms of the characters $\chi_\mR$, quadratic Casimirs $C_\mR$ and dimensions $d_\mR$ of the irreducible representations of the group, $\mR$, as
\be 
Z_{YM}(\kappa,c)=\sum\limits_\mR d_\mR\, \chi_{\mR}(c)\, e^{-\frac{ g_{YM}^2}{4} C_\mR\int d^2x \sqrt{g}\, \Phi^{-\frac{n+2}{2}}}\,.
\ee
The full path integral is then
\be \label{zc}
Z(c)=\sum\limits_\mR d_\mR \,\chi_{\mR}(c)\int\mathcal{D}g\mathcal{D}\Phi\, e^{-I_\mR}\,,
\ee
where the effective action $I_\mR$ takes the same form as \eqref{dilgra} but now with $U\to U_\mR$ such that
\begin{align}
    U_\mR(\Phi)=U(\Phi)+\frac{n(n-1)}{8 r_0^2 \phi_0^2} C_\mR\,\Phi^{-\frac{n+2}{2}}\,.
\end{align}
The last term modifies the solution by adding a `$\mR$otational' energy. In the limit near extremality the representation dependence only enters in a simple manner: as a small shift of the extremal energy due to the $SO(1,n-1)$ motion,
\be  
E_0 (\mR)\simeq E_0 +\delta E_0(\mR)\;,\qquad \delta E_0(\mR)=\frac{n-1}{4(n-2)}\frac{C_\mR}{\phi_b}\,,
\ee
and in the coupling of the effective JT theory, namely the dilaton (or area), $ \phi_0\to \phi_{0,\mR}$. The latter is a subleading modification, so, to first order we can set $\phi_{0,\mR}\simeq \phi_0$ and the extremal entropy is not modified. We must nevertheless retain $\delta E_0(\mR)$ since it is necessary to suppress the fluctuations of the motion in the group directions.

With these modifications, the sum over irreps in the partition function factorizes from the JT integral, and we obtain
\bea 
Z(c)=e^{S_0-\beta E_0}\,\left(\frac{\phi_b}{\beta}\right)^{3/2}\,e^{\frac{2\pi^2 \phi_b}{\beta}}\,\sum\limits_\mR d_\mR\,\chi_{\mR}(c) \, e^{-\beta\,\delta E_0 (\mR)}\,.
\eea
Our black holes are not rotating, so the characters are evaluated on the identity and $\chi_{\mR}(c)=d_\mR$. The sum over $\mR$ is a discrete one for compact groups. Instead, we have a non-compact group, so we must integrate
using the Plancherel measure  $\rho(\mR)$ for the irreps of $SO(1,n-1)$. 

For our purposes we do not require any details other than the temperature dependence, and this is easy to extract at low temperatures (see appendix~\ref{app:free}). Since $C_\mR\sim \lambda^2$ (with eigenvalue $\lambda$), the integral in this regime is dominated by the behavior at small $\lambda$, and for all $H_{n-1\geq 2}$ the Plancherel measures $\rho(\lambda)$ have the same behavior \cite{Camporesi:1990wm}
\begin{align}\label{rholambda}
    \rho(\lambda)\sim \lambda^{2}\,.
\end{align}
The integral over the group then gives
\begin{align}\label{betainhom}
    \int d\lambda\,\lambda^{2} e^{-\#\beta \lambda^2/\phi_b}\sim \left(\frac{\phi_b}{\beta}\right)^{3/2}\,,
\end{align}
which further suppresses the partition function at low temperatures.
Two comments are in order: (i) We have subtracted the constant part of the quadratic Casimir (e.g., $C(\lambda)-1/4 =\lambda^2$ for $SO(1,2)$, $n=3$). This subtraction has been argued in \cite{Iliesiu:2019xuh} for the Schwarzian $SO(2,1)$ modes, but we have not managed to justify it for the hyperboloid modes. If the constant term should be kept, then the gap would inhibit these fluctuations when $T\ll M_b$ and, simply, they would not contribute any additional factor like \eqref{betainhom}. (ii) The regularization of the hyperbolic volume $V_H$ breaks some of the symmetries. This makes the question of how they contribute to the partition function a subtle one. One might expect that a consistent procedure exists in which the regularization is only effectively done at the end of the computation and therefore \eqref{rholambda} holds. We will not attempt to fully solve these problems here, since the final result can only enhance (at least never reduce) the rate at which the partition function near extremality decreases towards zero. That is, the conclusion that the entanglement between the two sides vanishes as we approach the ground state is robust. 

\section{Disentangling de Sitter}\label{sec:dS}

Although we have focused on the entanglement of a CFT in Rindler space, it is known that a suitable Weyl transformation maps it to the conformally equivalent problem of the entanglement in $\mathbb{R}_t\times S^{n-1}$ across a partition into two hemispheres \cite{Czech:2012be}, or of a spherical entangling surface in flat space \cite{Casini:2011kv}. In a similar way, we can decrease the entanglement across the cosmological horizon of CFT states in de Sitter space.\footnote{This section was prompted by a discussion with Lenny Susskind.} As in \eqref{psibeta}, this is done by lowering the CFT temperature $T$ below the value $T_{dS}=1/2\pi$ of the unit-radius de Sitter universe. The Boulware-de Sitter state of the CFT is reached when $\beta\to\infty$.

To this purpose, we write the metric of the hyperboloid $H_{n-1}$ as \cite{Barbon:2012zv}
\begin{align}
    ds^2\left(H_{n-1}\right) = \frac1{1-\sigma^2}\left(\frac{d\sigma^2}{1-\sigma^2}+\sigma^2 d\Omega_{n-2}\right)\,,
\end{align}
with $0\leq \sigma<1$. The boundary, instead of the plane $\zeta=0$ of the Poincar\'e upper-half-space, is now the sphere $S^{n-2}$ at $\sigma\to 1$. With this expression the hyperbolic black hole solution \eqref{hyperbh} can be recast in the form
\begin{align}\label{hyperbh2}
    ds^2=\frac{r^2}{1-\sigma^2}\left( -f(r) (1-\sigma^2) dt^2 +\frac{d\sigma^2}{1-\sigma^2} +\sigma^2 d\Omega_{n-2} \right)+\frac{\ell^2}{r^2} \frac{dr^2}{f(r)}\,,
\end{align}
with $f(r)$ as in \eqref{fofr}. The boundary geometry at $r\to\infty$ is now in a conformal frame where the metric is that of dS$_{n}$,
\begin{align}\label{dsbdry}
    ds^2|_{\partial \mathcal{M}}= - (1-\sigma^2) dt^2 +\frac{d\sigma^2}{1-\sigma^2} +\sigma^2 d\Omega_{n-2}\,,
\end{align}
with a cosmological horizon at $\sigma=1$ such that $T_{dS}=1/2\pi$. When the bulk is a black hole with $\mu_0\leq \mu<0$, the temperature of the dual CFT is $T<T_{dS}$. Its entropy, regularized with a cutoff at $\sigma=1-\varepsilon$, is interpreted as the entanglement entropy of the CFT across the cosmological horizon. The analysis of the previous section implies that the Boulware-de Sitter state in the limit $\beta\to\infty$, dual to an extremal black hole, has large quantum fluctuations that bring the density of entangled states to zero.

\section{Outlook: Extreme quantum bridge demolition} \label{sec:discuss}

The apparent presence of a large entropy in non-supersymmetric extremal black holes has long been regarded as a puzzle. It takes an even more disconcerting guise in our setup, where it appears as a non-zero entanglement entropy of the Boulware product state $|0\rangle_L |0\rangle_R$. This would be a manifest inconsistency in AdS/CFT holography, and it demands a solution.

There is a sense in which all the entropies of non-BPS extremal black holes (at least in AdS) admit an interpretation of this kind. Whenever $T\neq 0$, the dual states are thermofield doubles of the theories on the disconnected asymptotic boundaries of the black hole. When $T\to 0$, one expects to recover the product state of the respective vacua, with vanishing entanglement entropy. 
In this article, we have placed the two CFTs side by side in Rindler space. Nevertheless, if $T\neq 2\pi$ they are actually disconnected: no CFT excitation can be sent from one side to the other. In the bulk, this would require sending a signal across a non-traversable Einstein-Rosen bridge. 

The puzzle we have described, and then resolved, is that in the limit $T\to 0$ this bridge, which becomes infinitely long, retains a finite width, i.e., finite area. As in the case of charged extremal black holes, what is missing here is the dominance of quantum effects down the long throat. There, quantum fluctuations of gravitational zero modes become strongly coupled and invalidate the semiclassical geometric description. The area of the bridge is no longer well-defined, so it is not a measure of the microscopic entanglement between the two sides. The gravitational path integral can controllably account for these quantum effects and yields a density of entangled states that vanishes when the energy above the ground state approaches zero.

The analysis involves a few subtleties that may be worth recounting. One of them concerns the instabilities of hyperbolic spaces to spontaneous nucleation of branes at low temperatures \cite{Barbon:2010us}. These are similar to issues about spontaneous superradiant decay or discharge in other non-BPS near-extremal black holes, which should be revisited taking proper account of quantum effects. The Rindler space interpretation also raises the issue of the need to keep the Rindler regulator  $\varepsilon$ finite. We can think of it as a simple proxy for a physical cutoff that makes manifest the difficulties with the holographic entanglement entropy. It is also needed to have a non-zero mass scale $M_b$ at which quantum effects become dominant, but it explicitly breaks symmetries of $H_{n-1}$. This may deserve closer attention, but, as we have seen, it does not modify our main conclusion.

Usually, quantum corrections are incorporated in holographic entanglement entropy via quantum extremal surfaces \cite{Engelhardt:2014gca}. Here, instead, we have employed the map between Rindler space and $\mathbb{R}\times H_{n-1}$ to compute, in the spirit of \cite{Casini:2011kv}, the entanglement entropy as a quantum thermal entropy. Then the Schwarzian theory gives the dominant quantum contribution to the entanglement entropy. Near $T=0$ its effect is to almost entirely cancel the classical area term, so it seems that one can no longer talk about a quantum extremal surface, because this requires a semiclassical geometry. It may be interesting to place the entanglement entropy at very low entanglement temperatures within the framework of \cite{Lewkowycz:2013nqa,Faulkner:2013ana}.

Perhaps the most surprising consequence is that the semiclassical spacetime born out of quantum entanglement can break down in situations where the geometry is weakly curved. As we have seen, infrared quantum gravitational fluctuations can become strong enough to bring about the demolition of geometric bridges. Reverting the process, can we see how a large spacetime gradually assembles from random matrices at the edge of the spectrum of extremal black holes?

\section*{Acknowledgments}

We are grateful to José Barbón, Alex Belin, Alejandro Cabo-Bizet, Horacio Casini, Rob Myers, Martín Sasieta, Lenny Susskind, Tadashi Takayanagi, Marija Tomašević and Joaquín Turiaci for conversations. RE is supported by MICINN grants PID2019-105614GB-C22 and PID2022-136224NB-C22, AGAUR grant 2017-SGR 754, and State Research Agency of MICINN through the ``Unit of Excellence María de Maeztu 2020-2023'' award to the Institute of Cosmos Sciences (CEX2019-000918-M). JMM is supported by CONICET, Argentina. RE would like to thank the Isaac Newton Institute for Mathematical Sciences for support during the programme ``Black holes: bridges between number theory and holographic quantum information'' when work on this paper (on the destruction of bridges) was undertaken, supported by EPSRC Grant Number EP/R014604/1.

\appendix

\section{Rindler entanglement for free conformal fields}
\label{app:free}

The states $|\Psi_\beta\rangle$ for free conformal field theories can be obtained by first solving for the thermal state of the field in $\mathbb{R}\times H_{n-1}$, which can be done with spectral methods, and then Weyl-transforming to Rindler space \cite{Candelas:1978gf,Emparan:1999gf}. We are interested in the behavior of the entropy for $\beta\gg 1$. For gauge fields, contact terms at the Rindler horizon can modify the numerical value of the entropy \cite{Eling:2013aqa}, but what will matter here is not its precise value, but only the generic behavior at low temperatures. 

\subsection{Leading divergent terms}

The energy of a free field of a given spin in $\mathbb{R}\times H_{n-1}$ at inverse (dimensionless) temperature $\beta$ takes, up to numerical factors, the form
\begin{align}
    E(\beta) \sim V_H\int_0^\infty d\lambda \,\frac{\lambda\,\rho(\lambda)}{e^{\beta\lambda}\mp 1}\,.
\end{align}
The Plancherel measure $\rho(\lambda)$ is obtained from the spectrum of the corresponding wave operator of the field in $H_{n-1}$ \cite{Camporesi:1990wm}. The $\mp$ sign in the denominator is for boson/fermion statistics.

The low-temperature behavior, $\beta\to\infty$, is controlled by the range of small $\lambda$ in the integrand, so that 
\begin{align}
     E(\beta) \sim V_H \frac{\rho(\lambda\sim\beta^{-1}\to 0)}{\beta^2}\,.
\end{align}
For conformal scalars in any dimension, $\rho(\lambda\to 0)\sim \lambda^2$. For spinors in any dimension, and for $p$-form gauge fields in $n=2p+2$ (such as four-dimensional Maxwell fields), $\rho(\lambda\to 0)\sim \lambda^0$. Therefore, whenever spinors or these $p$-form fields are present, they dominate the low-temperature behavior on the hyperboloid, with the energy vanishing as
\begin{align}\label{endens}
    E(\beta) \sim \frac{V_H}{\beta^2}\,,
\end{align}
and then the entropy
\begin{align}\label{sdens}
    S(\beta)=\int \beta\, dE \sim \frac{V_H}{\beta}\,.
\end{align}
If only conformal scalars are present, then the vanishing as $\beta\to\infty$ is faster, $E\sim \beta^{-4}$ and $S\sim \beta^{-3}$. Note that \cite{Almheiri:2016fws} has argued that the low-temperature behavior in \eqref{endens} and \eqref{sdens} is generically expected. 

The entanglement entropy in Rindler space \eqref{rindler} is the same as the thermal entropy in the hyperboloid \eqref{sdens}, only that the volume factor is reinterpreted: the large hyperbolic volume divergence becomes a short distance divergence near the Rindler horizon. Integrating over $\{\zeta,\mathbf{x}\}$ as in \eqref{VHreg} gives
\begin{align}
    S(T\ll 1) \sim \frac{A_{n-2}}{\varepsilon^{n-2}}T\,.
\end{align}

\subsection{Universal terms}

The leading divergent term of the entanglement entropy, although dependent on the cutoff, is enough to raise the puzzle that we address in this article, since it is present with any physical regularization. Nevertheless, it is interesting to also consider the better-defined `universal' terms in the entanglement entropy---universal because they are independent of the cutoff, although their values depend on the specific field content of the theory. Following \cite{Casini:2010kt}, here we will give results for massless scalar fields in even dimensions.

We start from the Renyi entropies defined as
\be 
S_\alpha=\frac{\log (\textrm{Tr}\rho^\alpha)}{1-\alpha}\,.
\ee
The conformal mapping to the hyperboloid allows us to compute these traces as thermal partition functions, so they can be written as
\be 
S_\alpha=\frac{1}{1-\alpha} \left(\log(Z(\alpha\beta))-\alpha\log(Z(\beta))      \right)\,.
\ee
Ref.~\cite{Casini:2010kt} used this expression to compute the Renyi entropies for a sphere of radius $R$ in even dimension $n$. The coefficients of the logarithmic divergence
\be 
S_\alpha=g_0^{(\alpha)}\log (\varepsilon/R)+\textrm{non-universal terms}
\ee
are cutoff-independent, i.e., universal. In the lowest three dimensions, they are
\bea \label{unirenyi}
g_0^{(\alpha)}&=& -\frac{\alpha+1}{6\alpha}\;,\qquad n=2\;,\nonumber\\
g_0^{(\alpha)}&=& \frac{(\alpha+1)(\alpha^2+1)}{360\, \alpha^3}\;,\qquad n=4\;,\nonumber\\
g_0^{(\alpha)}&=& -\frac{(\alpha+1)(3\alpha^2+1)(3\alpha^2+2)}{30240 \,\alpha^5}\;,\qquad n=6\;.
\eea
Since the index $\alpha$ can be understood as an effective modular inverse temperature, the Renyi entropies (functionals of an unnormalized density matrix $\rho^\alpha$) can be related to the entropy of the thermal density matrix that we are interested in\footnote{The holographic Renyi entropies for large Renyi index $\alpha$ were also studied in \cite{Belin:2013dva}, who considered a holographic theory with a low dimension scalar operator, leading to a different dominant phase at low energy. It would be interesting to analyze how the quantum effects we study affect the results of \cite{Belin:2013dva}.}. Defining 
\begin{align}
\tilde{\rho}=\frac{\rho^\alpha}{\textrm{Tr}(\rho^\alpha)}\,,
\end{align}
the relation is
\be 
S_\alpha(\tilde{\rho})=\alpha^2\partial_\alpha \left(\frac{\alpha-1}{\alpha}S_\alpha\right)\;.
\ee
The large $\alpha$ limit is equivalent to the large $\beta$ limit for the Rindler thermofield double. Using \eqref{unirenyi} we readily verify that
\be 
S_{\alpha\to\infty}(\tilde{\rho})=0\,,
\ee
as expected for the Boulware vacuum. 

This limit is not trivial. It requires that 
\begin{align}
    g_0^{(\alpha)}\;\propto\; 1+\frac1{\alpha} +O(1/\alpha^2)\,,
\end{align}
otherwise one finds non-zero answers in the limit. Notice also that the Renyi entropies do not vanish at large $\alpha$, where they are controlled by the smallest eigenvalue of the modular Hamiltonian (the modular vacuum). On the other hand, the entropy we are computing measures the entanglement degeneracy of this vacuum, which is expected to be zero on general grounds. 

The vanishing of the universal term is at odds with the holographic calculation from the RT formula. This can be written as
\be 
S_{RT}= s(\beta) V_{H}\,,
\ee
with $s(\beta)$ given in \eqref{sbh}. The divergences come from $V_H$ which, for comparison with the results above (see \cite{Casini:2010kt}), we expand in powers of $\varepsilon=R_s-r$, where $r$ is an infrared cutoff in the hyperboloid and $R_s$ is the radius of the boundary sphere. The expansion yields the holographic universal term as
\be \label{pruni}
S_{RT}= s(\beta)\,\frac{\Omega_{n-2}}{\left(2\pi\right)^n}\, q_0^{(n-1)} \log \left(\varepsilon/R_s\right)+\textrm{non-universal terms}\,,
\ee 
with
\be 
q_0^{(n-1)}\equiv\sum\limits_{j=0}^{n-2}(-1)^{n-1+j}\frac{(n+j-2)!}{(2j)!!(n-j-2)!j!}\;.
\ee
Since $s(\beta)$ remains non-zero as $\beta\to\infty$, the universal entanglement entropy term from the RT formula fails to vanish as it should in the Boulware vacuum. 

Large entangled degeneracies in the modular vacuum may exist only in supersymmetric quantum field theories. In the holographic context, supersymmetric black holes in AdS$_4$ with hyperbolic horizons and non-zero entropy have been studied in \cite{Caldarelli:1998hg,Cacciatori:2009iz,Halmagyi:2013sla,Gnecchi:2013mta,Benini:2016rke, Cabo-Bizet:2017jsl,Azzurli:2017kxo}.

\section{Coordinates}
\label{app:coords}

In Minkowski spacetime
\begin{align}
    ds^2=-du\,dv +d\mathbf{x}^2\,,
\end{align}
the left and right Rindler wedges are, respectively,
\begin{align}\label{rindcoords}
(u,v)=(\zeta\, e^{t},-\zeta\, e^{-t}) \qquad \textrm{and}\qquad (u,v)=(-\zeta\, e^{-t},\zeta\, e^{t}) \,,
\end{align}
with $\zeta>0$, $-\infty<t<\infty$.
 The transformation makes clear that, when $t$ is continued to imaginary time, regularity demands that it be identified with period $2\pi$, hence the temperature $T=1/2\pi$. 

The Rindler-AdS metric \eqref{rindlerads} with the Rindler space \eqref{rindler} at the boundary is obtained from the Poincar\'e-AdS metric
\begin{align}\label{pads}
    ds^2=\frac{\ell^2}{z^2}\left(dz^2 -du\,dv +d\mathbf{x}^2\right)
\end{align}
by changing $(u,v,z)\to (t,\zeta,r)$ as
\begin{align}
    u=-\zeta\,e^{-t}\,\sqrt{1-\frac{\ell^2}{r^2}}\,,\qquad v=\zeta\,e^{t}\,\sqrt{1-\frac{\ell^2}{r^2}}\,,\qquad z=\ell\,\frac{\zeta}{r}\,.
\end{align}
The transformation appropriate for the opposite wedge of Rindler-AdS is apparent from \eqref{rindcoords}.

\section{Holographic entanglement in two-dimensional Rindler space }
\label{app:2d}

Rindler-AdS$_3$ spacetime can be written as
\begin{align}
    ds^2&=-(r^2-\bar\mu)dt^2+\frac{\ell^2 dr^2}{r^2-\bar\mu}+r^2\frac{d\zeta^2}{\zeta^2}\nn\\
    &=\frac{r^2}{\zeta^2}\left[ -\left(1-\frac{\bar\mu}{r^2}\right) \zeta^2 dt^2 +d\zeta^2 
    \right]+\frac{\ell^2}{r^2} \frac{dr^2}{1-\frac{\bar\mu}{r^2}}\,.\label{rindlerads3}
\end{align}
This is the same as \eqref{hyperbh} with $n=2$, differing only by an inessential shift $\bar\mu=\mu+\ell^2$.
At the boundary, we find two-dimensional Rindler space, so these are the correct geometries to describe the Rindler entanglement of holographic CFTs in two dimensions.

If $\log\zeta$ were identified periodically, this would be a spinless BTZ black hole. Nevertheless, for the holographic Rindler interpretation we take $\zeta \in (0,\infty)$, resulting in AdS$_3$ in accelerated coordinates for all $\bar\mu>0$ (and not BTZ). The different values of $\bar\mu$ are merely due to a different normalization of the time coordinate $t$ in the metrics \eqref{rindlerads3}. This implies that the temperature of the horizon scales with $\bar\mu$ as
\begin{align}
    T=\frac{\sqrt{\bar\mu}}{2\pi\ell}\,.
\end{align}
The solution with $\bar\mu=0$ is the analog of the extremal black holes that we studied above. This solution is the Poincar\'e-AdS$_3$ metric (the `unwrapped massless BTZ').\footnote{The solutions with $-\ell^2\leq \bar\mu<0$ are not valid for our purposes (in the notation of \eqref{hyperbh} they have $-2\ell^2\leq \mu<-\ell^2$). If $\log\zeta$ were periodically identified, they would correspond to conical singularities or to global AdS$_3$. Without this identification, they have infinite conical excess angles.} We expect that it describes the dual of the Boulware vacuum in two dimensions.

Instead of the power-law divergence \eqref{VHreg} of $V_H$ we have a logarithmic one,
\begin{align}
    V_H=\int^L_\varepsilon \frac{d\zeta}{\zeta}=\log\frac{L}{\varepsilon}\,,
\end{align}
where in addition to $\varepsilon$ we have introduced an infrared cutoff $L$. The energy and entropy of the solution are
\begin{align}\label{ES2}
    E=\frac{V_H}{16\pi G}\frac{\bar\mu}{\ell}=\frac{2\pi^2}{M_b}T^2\,,\qquad S=\frac{V_H}{4G}\sqrt{\bar\mu}=\frac{4\pi^2}{M_b}T\,,
\end{align}
with 
\begin{align}
    M_b^{-1}=\frac{V_H\ell}{8\pi G}\,.
\end{align}

These features are easy to understand. What we have described is nothing but the analysis in \cite{Ryu:2006bv} of the holographic entanglement entropy of a two-dimensional CFT, only presented in the modular Rindler frame and using the modular temperature $T$. In contrast with $n>2$, here all the cases with $\bar\mu>0$ are equivalent.

Notice that, even if there is a mass scale $M_b$, there is no zero-temperature entropy $S_0$. 
So there is no entanglement entropy puzzle.
This does not mean that quantum effects are not important at low temperatures. For BTZ or whenever $M_b$ is non-zero, they must be, but they are not universally captured by the Schwarzian theory. To be clear, one can perform a dimensional reduction of three-dimensional gravity in the spinless sector, with an ansatz like \eqref{homog} with $n=2$, and find a two-dimensional JT theory with $\phi_0=0$. Its only dynamics is captured by a one-dimensional Schwarzian theory. This is valid even though we are not near extremality.
The classical solution of the Schwarzian \eqref{schwsol} reproduces the properties \eqref{ES2} for all values of $\bar\mu$. 

While the Schwarzian theory is thus a consistent truncation of the classical three-dimensional gravitational theory, its quantum fluctuations do not dominate the low-temperature regime (unlike in the extremal solutions with $n>2$ discussed in the main text). In contrast to the universal Schwarzian sector of CFTs and BTZ black holes near extremality \cite{Ghosh:2019rcj}, there is no such universality in the low-energy spectrum of three-dimensional quantum gravity or CFT with zero spin.

\bibliographystyle{utphys}
\bibliography{main}

\end{document}